\documentclass[a4paper]{article}

\usepackage{a4wide}
\usepackage{alloy}
\usepackage{tla}
\usepackage[T1]{fontenc}
\usepackage[scaled]{beramono}
\usepackage{graphicx}
\usepackage{subcaption}
\usepackage{amssymb}

\newcommand{\TLA}{$\mathsf{TLA}$}
\newcommand{\TLAplus}{$\mathsf{TLA}^+$}
\newcommand{\TLC}{$\mathsf{TLC}$}
\newcommand{\Alloy}{$\mathsf{Alloy}$}
\newcommand{\AlloyF}{$\mathsf{Alloy~5}$}
\newcommand{\AlloyS}{$\mathsf{Alloy~6}$}
\newcommand{\Analyzer}{$\mathsf{Analyzer}$}
\newcommand{\Electrum}{$\mathsf{Electrum}$}
\newcommand{\nusmv}{$\mathsf{NuSMV}$}
\newcommand{\nuxmv}{$\mathsf{nuXmv}$}
\newcommand{\spin}{$\mathsf{SPIN}$}
\newcommand{\ProB}{$\mathsf{ProB}$}
\newcommand{\B}{$\mathsf{B}$}
\newcommand{\FOL}{$\mathsf{FOL}$}
\newcommand{\LTL}{$\mathsf{LTL}$}
\newcommand{\DSL}{$\mathsf{DSL}$}
\renewcommand{\LaTeX}{$\mathsf{LaTeX}$}

\begin{document}

\title{Designing Software with Complex Configurations}
\author{Alcino Cunha}
\date{July 2024}

\maketitle

\section{Introduction} 

The design phase is key to achieve high-quality software systems. The term \emph{design} can mean different things to different people, but for me it is just the synonym of an \emph{abstraction} or \emph{model} of the system to be built. Different levels of formality can be used to define such models, but to properly reason about the desired properties of the system we need rigorous mathematical models. These are particularly relevant if the system is mission- or life-critical, and computer science has a long tradition of applying so called \emph{formal methods} in the design of those kinds of software systems.
However, as I will show later, some rigorous notations and tools for describing and analyzing software designs are now so simple and usable, that there is little excuse to not apply them to the design of most software, critical or not. Maybe that could help prevent many of the misconceptions and flaws that plague software applications nowadays.

Software systems typically operate in a given \emph{configuration}. By configuration, I mean a set of parameters or conditions that affects the behavior of the system and whose value is assumed to be constant for \emph{a while}: in reconfigurable systems this \emph{a while} can be a (reasonable) time interval, but in many cases it is the whole life-span of the system, once configured and deployed. 

Just to make the term \emph{configuration} more precise, here are some examples of configurations in software systems with different granularity: in a \emph{software product line} designed against a particular \emph{feature model}, the configuration could be the set of features that will be selected in a given deployment; in the software controlling a railway interlocking system, the configuration could be the specific railway network (and signaling layout); in a replicated database, the configuration includes at least the number of replicas; in a distributed protocol, the configuration could be the network topology where the protocol will run; finally, in a concurrent algorithm, the configuration could be just the number of executing processes. 

Some of these configuration are extremely simple, for example the number of processes in a concurrent algorithm, but others are quite complex, for example the railway network in interlocking systems. Designing systems to work with such complex configurations is not trivial, because ideally we should ensure that the expected properties are satisfied whatever the configuration. And sometimes it is not even trivial to  enumerate all possible configurations, for example, enumerating all valid feature combinations of a feature model.

In this paper I will discuss \emph{if} and \emph{how} existing formal methods can be applied to the design of software with such complex configurations. I also intend to show that some formal methods are already cost-effective enough to be applied in the design of any software system, not just critical ones.
For that reason, I will focus on methods that can be applied by any software engineer with the standard background on logic and discrete math, and will immediately rule out many \emph{heavyweight} formal methods, namely those focused on obtaining full proofs (aka theorem provers) and that usually require expert user input.   Instead, I will focus on \emph{lightweight} formal methods, that provide automatic analyses, with the tradeoff that in many cases they will only be able to achieve partial proofs. Notice that these are still vastly superior to testing: first, many of the systems nowadays are distributed or concurrent, and bugs in those systems are typically due to very sporadic race conditions that are almost always impossible to catch with normal testing; second, even in a full deterministic system, they will be able to verify properties for all possible configurations up to a given size, as opposed to testing just a few configurations. 

To make the presentation more concrete I will use as running example the \emph{Echo} distributed protocol first proposed by Chang~\cite{Chang82}. This protocol aims to define a \emph{spanning tree} in an arbitrary connected network of nodes  with a distinguished \emph{initiator}. Such \emph{spanning tree} could be used, for example, to define routing tables for posterior communication. Chang describes the protocol roughly as follows.
The initiator starts by sending an \emph{explorer} message to all its neighbors. If a node receives an explorer and it is the first to arrive at the node, mark the sender as its \emph{parent} and send an explorer to all neighbors except the parent. 
If a node receives an explorer but it is not the first to arrive or there are no neighbors except the sender, send an \emph{echo} message back to the sender.
If the received message is an echo, register that an echo as been received from the sender, and if echos from all neighbors (to which explorers were sent) have arrived send an echo to the parent, unless the node is the initiator, in which case the protocol is finished.
The main properties of interest in this protocol are (partial) \emph{correctness} -- when the protocol finishes the parent relationship indeed forms a spanning tree rooted at the initiator -- and \emph{termination} -- the protocol eventually finishes. Recall that these should hold for every possible configuration, in this case, for every possible connected network and initiator.

\section{Software design with {\TLAplus}}

Being our example a distributed protocol, we cannot avoid starting the discussion with \TLAplus~\cite{Lamport02}, the language introduced by Leslie Lamport precisely to design distributed and concurrent algorithms. \TLAplus\ comes with a powerful model-checker -- \TLC\ -- and has been applied in the design of many real software systems, for example by Amazon~\cite{Newcombe14}. \TLAplus\ is based on the \emph{Temporal Logic of Actions} (\TLA)~\cite{Lamport94}, essentially a combination of \emph{First-Order Logic} (\FOL) and \emph{Linear Temporal Logic} (\LTL). Before \TLA, Amir Pnueli had already showed how \LTL\ could be used to reason about models of concurrent programs~\cite{Pnueli77}. The models in question were simple state transition systems, one of the quintessential models of computation and the modeling formalism used by most lightweight formal methods, namely model-checkers. The main novelty of \TLA\ was the introduction of \emph{actions}, formulas with primed and unprimed variables, which allow \TLA\ to be used not only to formalize the properties of the system, but also the model of the system itself, without the need for a specific \emph{Domain Specific Language} (\DSL), as is the case with most model-checkers. Unprimed variables are evaluated in the pre-state of a transition, while primed variables are evaluated in the post-state, and thus actions can be used to succinctly formalize the transition relation of a state transition system. Not having to learn two separate formal languages is obviously a plus for new users, and probably one of the reasons for the popularity of \TLAplus. Another advantage is that using plain logic to formalize the model is more adequate for the level of abstraction desired at the design phase, unlike some modeling {\DSL}s that tend to look a lot like a programming language. And as Leslie Lamport says~\cite{Lamport18} ``if you’re not writing a program, don’t use a programming language''!

A good thing about \TLAplus\ is that it clearly distinguishes the configuration of the system from its mutable state. The rigid variables that define the former are declared with keyword \t{CONSTANTS} while the flexible variables that define the latter are declared with \t{VARIABLES}.

\begin{tla}
CONSTANTS Node, Initiator, adj
VARIABLES parent, received, inbox    
\end{tla}

In our example, the protocol configuration will be specified by three constants: \t{Node}, the nodes of the network; \t{Initiator}, the initiator node; and \t{adj}, the neighbor relationship that defines the network topology. For the state we have three variables, that will store for each node: its \t{parent}, if any; the set of neighbors from which echos have already been \t{received};  and the set of unprocessed messages in the respective \t{inbox}. This protocol should work correctly even if messages are delivered out of order, something we will abstract by having the inboxes contain sets of messages and letting nodes pick any of them to process at each time.

\TLAplus\ is an untyped language and variables can take any value from standard mathematical types such as \emph{booleans}, \emph{numbers}, \emph{sets}, or \emph{functions}. By encoding them as functions, \TLAplus\ also supports \emph{records} (functions from field names to the respective values) and \emph{sequences} (functions from indexes to values). In our example, \t{Node} will be a set of node identifiers, \t{Initiator} a single node, and the remaining constants and variables will be functions associating each \t{Node} with the respective information. To manipulate expressions of each type, \TLAplus\ provides specific syntax and operators, sometimes resorting the well-known \LaTeX\ syntax (for example for set operators). In this document, the latter will be rendered with the standard mathematical notation, while the standard ASCII notation is used for the remaining. I believe that the syntax and semantics of most operators is trivial to understand for most readers with the usual background in logic and discrete math, and so I will use them without further explanation. 

After declaring the constants, we can specify the assumptions about the protocol configuration using the \t{ASSUME} keyword. 
\begin{tla}
reachable(n) ==
    LET aux[i \in Nat] == 
        IF i = 1 THEN adj[n]
        ELSE aux[i-1] \cup { x \in Node : \E y \in aux[i-1] : x \in adj[y] }
    IN  aux[Cardinality(Node)]

ASSUME /\ Initiator \in Node
       /\ adj \in [Node -> SUBSET Node]
       \* no self loops
       /\ \A n \in Node : n \notin adj[n]
       \* undirected graph
       /\ \A x,y \in Node : y \in adj[x] <=> x \in adj[y]
       \* all nodes reachable from initiator
       /\ Node \ {Initiator} \subseteq reachable(Initiator)
\end{tla}

Most of the assumptions are rather trivial to specify: the \t{Initiator} must be one of the nodes; the \t{adj} relation is a function from \a{Node} to subsets of \a{Node} (expression \t{[A -> B]} denotes the set of all functions from \t{A} to \t{B}, and \t{SUBSET A} the power-set of \t{A}); the adjacency relation does not contain self loops; and is also symmetric (the network is an undirected graph). However, the assumption that the network is connected is a bit more tricky to formalize. Since \TLAplus\ does not provide native closure operators, we must define the set of nodes reachable from any node using recursion: definition \t{reacheable(n)} first defines recursively an auxiliary function \t{aux} that computes the set of nodes reachable from \t{n} in at most \t{i} steps, and then determines the set of nodes that that are reachable from \t{n} in at most \t{Cardinality(Node)} steps using that function. Because of this definition we need to import the standard modules \t{Naturals} (to use the arithmetic operators) and \a{FiniteSets} (to use the \t{Cardinality} operator), using the \t{EXTENDS} keyword.

\begin{tla}
EXTENDS Naturals, FiniteSets
\end{tla}
    
Since the language is untyped it is easy to make mistakes when manipulating variables. It is customary to include in a \TLAplus\ specification a predicate that checks if the state variables contain values of the appropriate type. This predicate is frequently named \t{TypeOK} and can later be model-checked to verify its invariance. Although a poor substitute for static type checking, this can help find some typing errors, but also serves as documentation and it is good practice to declare it early in the specification. In our example we have the following \t{TypeOK} predicate.

\begin{tla}
CONSTANT None
ASSUME None \notin Node

TypeOK == /\ parent \in [Node -> (Node \cup {None})]
          /\ received \in [Node -> SUBSET Node]
          /\ inbox \in [Node -> SUBSET [from : Node, type : {"Explorer","Echo"}]]
\end{tla}
        
In \TLAplus\ a partial function like  \t{parent} can be defined by a normal function, using a distinguished constant \a{None} as result when there is no value associated with a particular domain element.
The messages sent in this protocol will be represented by records with a \t{from} and a \t{type} field. The former will contain a node identifier and the latter a string identifying one of the two possibles types of messages (in \TLAplus, strings are uninterpreted literals, supporting only equality). In the predicate checking the type of \t{inbox}, expression \t|[from : Node, type : {"Explorer","Echo"}]| denotes the set of all records where field \t{from} is a node identifier and \t{type} is one of the two strings.

With \TLA, a transition system can implicitly be specified with a temporal formula that specifies which are its valid behaviors (or traces), being a behavior an infinite sequence of states starting in an initial state and where each pair of consecutive states respects the transition relation. This temporal formula has the shape $I \wedge \square [A]_t$, where $I$ is a state predicate that specifies which are the valid initial states, $\square$ is the standard \emph{always} \LTL\ operator that enforces the enclosed formula to be true in all states of a behavior, $A$ is an \emph{action} that specifies if two consecutive states satisfy the transition relation, and $[A]_t \doteq A \vee t' = t$ is a formula that holds if $A$ holds or $t$ has the same value in the pre- and post-state. Note that, whatever $t$, $[A]_t$ will always allow \emph{stuttering} steps, where all variables keep the same value. If $t$ is a constant then $t' = t$ is a tautology and the system will be able to evolve freely, with all variables potentially changing their value in an arbitrary way. However, usually $t$ is a sequence with all state variables, and in this case formula $\square [A]_t$ will specify that it is always the case that the system performs an $A$ step or stutters. In \TLA, actions cannot be used freely in formulas, and can only be used inside the operator $[\,\cdot\,]_t$. This is to ensure that all \TLA\ formulas are invariant under stuttering, and thus will not have their validity affected by refinement (for example, adding extra variables to a system). Lamport claims that this is a fundamental property of a logic to specify systems, usually illustrating his claim with a very simple example of a clock specification. A clock showing only the hours can be specified with a single $h$ variable as follows.
\begin{displaymath}
    0 \le h < 24 \wedge \square [h' = (h + 1) \mathbin{\%} 24]_h
\end{displaymath}
Initially we have $h$ with an arbitrary hour and then the system evolves by changing the hour as expected. If this specification did not allow for stuttering steps, then a clock with hours and minutes would not satisfy it, because the hour would be forced to always keep changing.

In a larger system with many different events it is usual for the action $A$ in the specification $I \wedge \square [A]_t$ to be a disjunction of more specific actions specifying the different events. That is the case of the specification of this protocol, where the action that specifies the transition (or next state) relation (denoted \t{Next} below) is a disjunction of all possible events that can occur in all possible nodes. In this case, two possible events can occur at each node, namely receive an explorer or an echo message.
\begin{tla}
vars == << received, parent, inbox >>
Next == \E n \in Node : receiveExplorer(n) \/ receiveEcho(n)
Spec == Init /\ [][Next]_vars
\end{tla}
Here, \t{vars} is a sequence with all the state variables that should remain unchanged in a stuttering step.
The initial state predicate can be specified as follows.
\begin{tla}
msg(n,t) == [ from |-> n, type |-> t ]

Init == \* no parents
        /\ parent = [ n \in Node |-> None ]
        \* no received echos
        /\ received = [ n \in Node |-> {} ]
        \* initiator starts by sending explorers to all neighbors
        /\ inbox = [ n \in Node |-> IF n \in adj[Initiator] 
                                    THEN {msg(Initiator,"Explorer")} 
                                    ELSE {} ]
\end{tla}
The initial value of the three variable functions is defined by comprehension. Since the protocol starts with the initiator sending a ping to all its neighbors, the initial value of function \t{inbox} is specified accordingly by comprehension.

To give an example of how to use actions to specify an event, consider the definition of \t{receiveExplorer(n)} that specifies the event of node \t{n} receiving and processing an explorer message.
\begin{tla}
receiveExplorer(n) == \E m \in inbox[n] :
    /\ m.type = "Explorer"
    /\ IF parent[n] = None 
       THEN parent' = [ parent EXCEPT ![n] = m.from ]
       ELSE UNCHANGED << parent >>
    /\ IF parent[n] = None /\ adj[n] \ { m.from } # {}
       THEN inbox' = [ a \in Node |-> 
                       CASE a \in adj[n] \ {m.from} -> inbox[a] \cup {msg(n,"Explorer")}
                         [] a = n  -> inbox[a] \ {m}
                         [] OTHER  -> inbox[a] 
                     ]
       ELSE inbox' = [ inbox EXCEPT ![n] = @ \ {m}, ![m.from] = @ \cup {msg(n,"Echo")} ]
    /\ UNCHANGED << received >>            
\end{tla}
The specification of an event usually comprises three kinds of conjuncts: \emph{guards}, formulas specifying when can the event occur; \emph{effects}, that specify which variables change and how do they change; and \emph{frame conditions}, that specify which variables do not change. In this case, the guard states that the event can occur if there exists an explorer message in the inbox of \t{n}. When the event occurs there is an effect on two of the state variables: if it is the first explorer to be received, the sender is set as the \t{parent} of the node (the \t{EXCEPT} keyword is used to define a function that is equal to another except for a given domain element, with \t{@} denoting its old value); if it is the first explorer and there are more neighbors besides the sender an explorer is sent to all of those; otherwise an echo is sent back; in any of these cases the received message is deleted from the \t{inbox}.
Finally, a frame condition specifies that this event does not change the value of \t{received}. The keyword \t{UNCHANGED} is used to specify that a variable or a tuple of variables does not change its value.

Before verifying the expected properties it is important to validate the protocol specification. \TLAplus\ does not have an interactive simulator that allows the user to validate a specification. The recommended validation technique is to write false properties that force the \TLC\ model-checker to provide as counter-examples some expected scenarios. For example, we could specify a property stating that it is impossible for the protocol to finish, and, if the protocol is well-specified, a counter-example for this property will be a minimal sequence of steps where the protocol runs to completion.

\begin{tla}
Finish == received[Initiator] = adj[Initiator]
MinimalRun == [](~Finish)
\end{tla}

\begin{figure}[t]
    \centering
    \includegraphics[width=12cm]{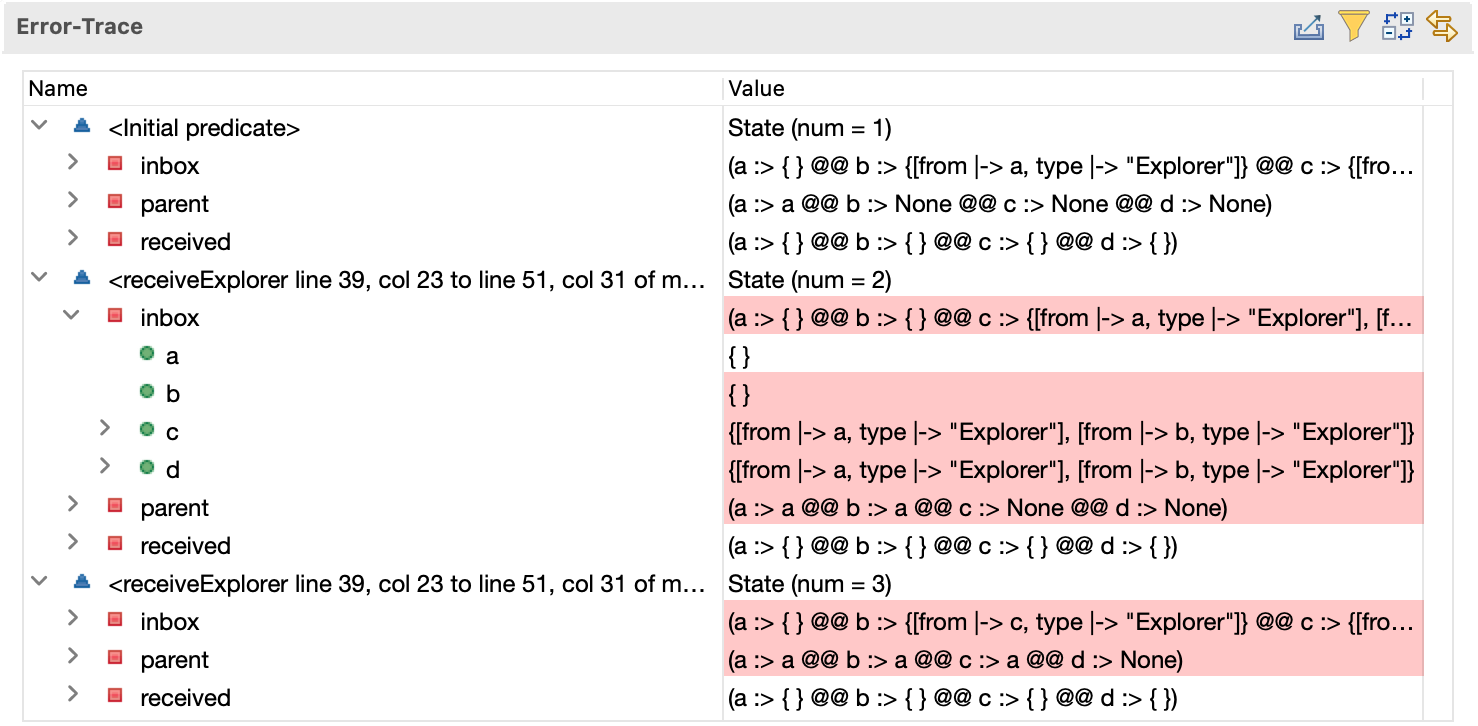}
    \caption{Minimal run in a complete network with 4 nodes}
    \label{fig:complete4tla}
\end{figure}

To see such a scenario for a complete network with 4 nodes (\t{a},\t{b},\t{c}, and \t{d}), we first need to instantiate the constants \t{Node}, \t{Initiator}, and \t{adj} accordingly. Then we can ask \TLC\ to verify property \t{MinimalRun}, and a trace with 19 states is returned. Counter-examples are displayed in a textual format, highlighting the variables that change value at each state, as seen in Figure~\ref{fig:complete4tla} (showing only the first 3 states). It is possible to hide or expand the value of a variable, as done for the \t{inbox} in the second state. In this trace the first event consisted of node \t{b} reading the explorer message sent by node \t{a} (the initiator): after this event, its \t{inbox} is empty and the inboxes of nodes \t{c} and \t{d} contain explorers sent both by the initiator \t{a} and by \t{b}. 

Having validated our protocol specification, we can proceed to the verification of the two expected properties, correctness and termination. The former is an example of a safety property, and can be specified as follows.

\begin{tla}
SpanningTree == \A n \in Node \ {Initiator} : Initiator \in ancestors(n)
Correctness == [] (Finish => SpanningTree)
\end{tla}
The auxiliary definition \t{SpanningTree} states that all nodes must have the initiator as an ancestor. To determine the ancestors, again we need to resort to a recursive (omitted) definition similar to the one that was defined for computing the reachable nodes.

Since the specification of the protocol admits stuttering steps, to verify termination we need to impose fairness restrictions that force the system to make progress. In this case it suffices to require weak fairness for the \a{Next} transition relation, meaning that while there is at least a node with unprocessed messages the algorithm cannot stutter forever. \TLAplus\ has specific operators to specify fairness conditions (\t{WF} for weak fairness and \t{SF} for strong fairness), but they can be specified directly using the temporal connectives and the \t{ENABLED} operator (that checks if an action is enabled in a state). The termination property can thus be specified as follows.

\begin{tla}
Termination == WF_vars(Next) => <> Finish           
\end{tla}

\TLC\ is extremely efficient, in particular when verifying invariants. For the configuration consisting of a complete network with 4 nodes both properties are checked in around 2s.
\TLC\ implements an explicit model-checking technique that traverses the state space, but takes advantage of multi-core processing to parallelize the search and uses hash tables to summarize states and reduce the memory footprint. Due to the need to explicitly compute the next states, \TLC\ cannot however accept any \TLAplus\ specification. In particular, the next state relation must be a disjunction of actions, each a conjunction of ``assignments'': the new value of a (primed) variable can only be specified with an equality (deterministic assignment) or an inclusion (non-deterministic assignment), and the assigned expression must only use variables assigned beforehand. Although these restrictions can make the specification of some systems a bit more difficult, when specifying concurrent and distributed algorithms they are rarely a limitation.

However, for a complete network with 5 nodes, \TLC\ already takes more than 4m just to check correctness and around 20m to check termination, even after turning on symmetry breaking to avoid exploring states that are isomorphic up to a permutation of the node identifiers. For example, in the complete network with 5 nodes with initiator \t{a} it is unnecessary to explore both states where \t{b} is the first to read the explorer or \t{c} is the first to read the explorer. By assuming the set \a{Node} is symmetrical only one of those will be explored. In this particular configuration \TLC\ has to explore around 3 million truly distinct states.

But the main problem with this \TLAplus\ specification is that \TLC\ can only be used to check one configuration at a time. With up to 4 nodes there are 16 truly different network configurations, and to check them all the user must manually configure \TLC\ with 16 different instantiations of the declared constants. Moreover, it is not at all easy to manually enumerate the 16 truly different (non-symmetrical) networks with up to 4 nodes (we leave that as an exercise to the reader). This strategy will obviously not scale if we want to verify the protocol with a higher number of nodes and there is a good chance that a faulty configuration can be missed.

Fortunately, it rather easy to adapt a \TLAplus\ specification to check all configurations. The basic idea is to change the declared constants to flexible variables, move the assumptions to constraints on the initial state, and add frame conditions to all events to force those configuration variables to remain constant in all behaviors. With this change every configuration will give rise to a different initial state to be explored by \TLC. As mentioned above, \TLC\ requires every variable to have its value defined either with an equality (corresponding to an assignment) or an membership test (corresponding to a non-deterministic assignment). In our case, this means that we have to add an extra constraint to \t{Init} to define the initial value of set \t{Node}. Since we want to explore all networks with up to a given number of nodes, we will add a constant \t{Univ} with all possible node identifiers and then assign to \t{Node} an arbitrary subset of \t{Univ}. Our specification will look roughly as follows.

\begin{tla}
EXTENDS Naturals, FiniteSets

CONSTANT Univ
VARIABLES received, parent, inbox, Node, Initiator, adj    

...

Init == /\ Node \in SUBSET(Univ)
        /\ Initiator \in Node
        /\ adj \in [Node -> SUBSET Node]
        \* no self loops
        /\ \A n \in Node : n \notin adj[n]
        \* undirected graph
        /\ \A x,y \in Node : y \in adj[x] <=> x \in adj[y]
        \* all nodes reachable from initiator
        /\ Node \ {Initiator} \subseteq reachable(Initiator)
        /\ ...

config == << Node, Initiator, adj >>    
receiveExplorer(n) == \exists m \in inbox[n] :
    /\ ...            
    /\ UNCHANGED config     

...
\end{tla}

\begin{figure}
    \centering
    \includegraphics[width=5cm]{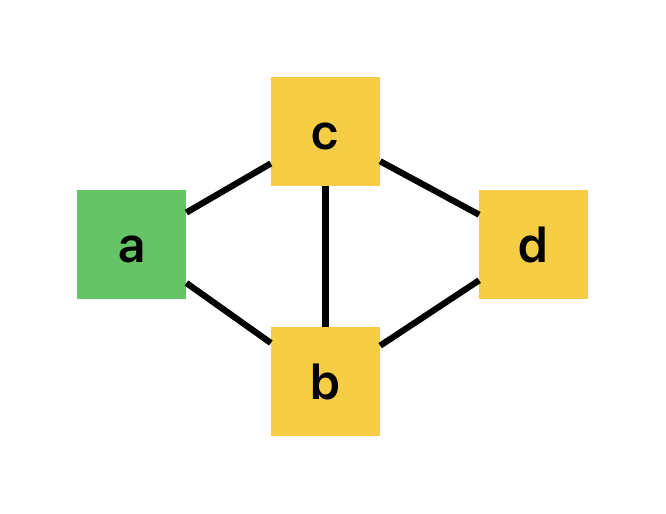}
    \caption{Buggy configuration}
    \label{fig:bad4}
\end{figure}

Unfortunately, when checking the expected properties for a universe of 4 possible node identifiers (\t{a}, \t{b}, \t{c}, and \t{d}) we immediately get a counter-example to property \t{Correctness} for the following configuration, corresponding to the network in Figure~\ref{fig:bad4} (the initiator is shown in green).
\begin{tla}
Node = {a,b,c,d}
Initiator = a
adj = a :> {b, c} @@ b :> {a, c, d} @@ c :> {a, b, d} @@ d :> {b, c}    
\end{tla}

In fact, out of the 16 non-symmetrical configurations only 2 of them are buggy, the other one being similar to the one on Figure~\ref{fig:bad4}, but with the edge from \t{d} to \t{c} removed. Note that, if the initiator was node \t{b} (or symmetrically \t{c}) then the protocol would work fine. Without an exhaustive verification of all configurations I recon it would be very unlikely that an user would choose to verify the protocol for these two configurations and the bug would most likely go unmissed.

So, is Chang's protocol really faulty? Not really... Somewhere in the paper, when presenting general definitions it is mentioned that the initiator should be considered visited \emph{a priori}. In the protocol description it is not clarified how this should be implemented, but a reasonable assumption could be to consider the initiator to already have a parent at the beginning. This can be specified as follows in the \t{Init} predicate (we will assume the initiator is its own parent).

\begin{tla}
Init == /\ ...
        \* initiator is explored a priori
        /\ parent = [n \in Node |-> IF n = Initiator THEN Initiator ELSE None ]
        /\ ...
\end{tla}

Checking our protocol again for all networks up to 4 nodes now yields no counter-example, and verification takes only a few seconds (to check both \t{Correctness} and \t{Termination}). Checking \t{Correctness} for all networks up to 5 nodes takes around 13m. This is not surprising as a single configuration takes around 4m. Checking \t{Termination} takes a bit over 1h.

\begin{table}
    \centering
    \begin{tabular}{c|c|c|c}
        \textbf{Nodes} & \textbf{Candidates} & \textbf{Non-symmetric} & \textbf{Symmetric}\\
        \hline
        $1 .. 3$ & 1570 & 21 & 5 \\
        $1 .. 4$ & 263714 & 216 & 16 \\
        $1 .. 5$ & 168035874 & 4545 & 74 \\
        $1 .. 6$ & 412484896290 & ? & 531         
    \end{tabular}
    \caption{Possible configurations}
    \label{tab:configurations}
\end{table}

However, if we try to check the protocol for networks with up to 6 nodes, we get stuck: after one hour \TLC\ still does not even finish to enumerate the possible initial states (corresponding to the different configurations). As we can see in Table~\ref{tab:configurations}, the problem is that the number of non-symmetric configuration is just a tiny fraction of all the possible valuations of the configuration variables. When dealing with non-deterministic assignments, either in initial states or in the effect events, \TLC\ just enumerates explicitly all possible candidates until finding those that satisfy the required conditions. Moreover, unlike in the state graph exploration, this enumeration is not parallelized. It is surprising that it can even finish to enumerate all the 74 non-symmetrical configurations for networks with up to 5 nodes in less than 1m (a testament to \TLC\ efficiency), but obviously this technique cannot scale up to more complex configurations. For networks with up to 6 nodes, this is really just like ``looking for a needle in a haystack''.
This kind of state-explosion was precisely the motivation for the development of symbolic model-finding and model-checking techniques that do not need to explicitly enumerate the state space.

\section{Structural design with \Alloy}

\Alloy~\cite{Jackson16} is a language and tool, proposed by Daniel Jackson at MIT, that was designed to excel at describing and analyzing different kinds of structures in software systems. Its \Analyzer\ implements precisely a symbolic model-finding technique (based on SAT solvers), that is particularly good at ``looking for a needle in a haystack'' (in particular, I used it to count the number of non-symmetric configurations in our running example, presented in Table~\ref{tab:configurations}). It allows the user to check expected properties, but unlike most of the formal design tools it also allows the user to  explore different scenarios without verifying properties, just by asking for examples of structures satisfying the specified requirements (instances, in \Alloy\ parlance). Both counter-examples and instances can be depicted graphically, with customized themes, which eases a lot the comprehension of complex structures. The \Analyzer\ also supports iteration, allowing the user to ask for different counter-examples or different instances. These features are extremely useful in helping the user elicit requirements or validate a specification.

But another key aspect of \Alloy\ is that its language was designed for abstraction, which is \emph{the crux} of formal software design. In particular, all structures have to be described using the single unified concept of mathematical relation. 
At first, this \Alloy\ motto that ``everything is a relation'' can be quite puzzling for users, but ends up forcing them to really think differently from programming. Even \TLAplus\ has several (mathematical) types that kind of resemble the types existing in many programming languages, so with \Alloy\ users are further coerced into the ``if you’re not writing a program, don’t use a programming language'' mindset recommended by Leslie Lamport for the design phase.

Most computer science graduates should be well familiar with the concept of binary relation (a set of pairs). Binary relations can be used, for example, to model functions or partial functions, by restricting each element in the domain to be related to exactly one or at most one element in the range, respectively. In \Alloy, relations can be of arbitrary arity, but uniform, meaning all tuples of elements contained in a relation should have the same length. Unary relations can thus be used to represent sets, and singleton unary relations can be used to denote specific elements of the domain. 

In \Alloy\ sets are known as \emph{signatures} and the elements that inhabit them as \emph{atoms}. The \a{sig} keyword can be used to declare signatures. Signature declarations can have a multiplicity attached to restrict the number of atoms they can contain. A signature can also \emph{extend} another signature, meaning the former is a subset of the later and disjoint from every other extension. A \emph{top-level} signature is one that does not extend another signature. This hierarchy is exploited in the type-system of \Alloy\ to find \emph{irrelevance} errors, expressions that always denote empty relations, and that are most likely specification mistakes~\cite{typesystem}.
Relations of arity higher than one are known as \emph{fields} and are declared inside the domain signature, with an optional multiplicity attached to the range. For example, to specify the configurations of the Echo protocol in \Alloy\ we could start by declaring the following signatures and fields.
\begin{alloy}
sig Node {
    adj : set Node
}
one sig Initiator extends Node {}
\end{alloy}    
Signature \a{Node} will contain the nodes of the network. To distinguish a specific node as the \a{Initiator} we declare a singleton signature extending \a{Node}. The neighbors are represented by the binary relation \a{adj} that associates each node with an arbitrary number of nodes.

Assumptions are specified inside \emph{facts}. These are specified with \emph{Relational Logic}, an extension of \FOL\ that, besides the usual quantifiers (\a{all} and \a{some} in \Alloy\ syntax) and membership and equality atomic formulas (\a{in} and \a{=}), supports a few derived operators that ease the specification of constraints, as well as closure operators that cannot be expressed in \FOL. Due to the ``everything is a relation'' motto, the \Alloy\ syntax is quite small and also has a very simple semantics -- the full list of operators is presented in Figure~\ref{fig:relational}. Note that set operators, like union or intersection, can be applied to relations of arbitrary arity. For example, operator \a{in}, that determines if a relation is a subset or equal to another relation, can also be used to check membership of an element to a set, since the former will be denoted by a singleton unary relation and the latter by a unary relation (for example, \a{Initiator in Node} is a possible atomic formula that holds trivially in our model).

The most used \Alloy\ operator is \emph{dot join}, that allows us to navigate through a relation to obtain related elements. It can by applied to any pair of relational expressions, as long as the sum of their arities is greater than two. The semantics of this operator is as follows.
\begin{displaymath}
    \begin{array}{c}
        R \mathbin{\ma{.}} S 
        \equiv 
        \{ (a_1,\ldots,a_{n-1},b_2,\ldots,b_m) \mid (a_1,\ldots,a_n) \in R \wedge (b_1,\ldots,b_m) \in S \wedge a_n = b_1\}        
    \end{array}
\end{displaymath}
Essentially, it concatenates all tuples from $R$ with all tuples from $S$ that have the same last and first atom, respectively, with the nuance that the matching atom is dropped. Since ``everything is a relation'' this operator has multiple uses. For example, if $f$ and $g$ are binary relations that represent functions, $f \mathbin{\ma{.}} g$ is the same as $g \circ f$ (function composition), and if $x$ is a singleton unary relation denoting a particular atom of the domain and $f$ is a binary relation denoting a function, $x \mathbin{\ma{.}} f$ is the same as $f(x)$ (function application) and  $f \mathbin{\ma{.}} x$ is the same as $f^{-1}(x)$ (inverse application). For example, \a{Initiator.adj} is the set of neighbors of the initiator, \a{adj.Initiator} is the set of nodes that have the initiator as neighbor, and \a{initiator.Node} is the set of nodes that have some neighbor.

\begin{figure}[t]
    \centering
    \begin{subfigure}[b]{0.35\linewidth}
        \begin{tabular}{cl}
            \ma{+} & union\\
            \ma{&} & intersection\\
            \ma{-} & difference\\
            \ma{->} & cartesian product\\
            \ma{<:} & domain restriction\\
            \ma{:>} & range restriction\\
            \ma{.} & dot join \\
            \ma{in} & subset or equal\\
            \ma{=} & equality
        \end{tabular}
        \caption{Binary operators}            
    \end{subfigure}    
    \begin{subfigure}[b]{0.35\linewidth}
        \begin{tabular}{cl}
            \ma{\~} & converse \\
            \ma{^} & transitive closure \\
            \ma{*} & reflexive closure\\
            \ma{no} & nothing\\
            \ma{some} & at least one\\
            \ma{lone} & at most one\\
            \ma{one} & exactly one\\
        \end{tabular}
        \caption{Unary operators}            
    \end{subfigure}    
    \begin{subfigure}[b]{0.28\linewidth}
        \begin{tabular}{cl}
            \ma{none} & empty set \\
            \ma{univ} & universe \\
            \ma{iden} & identity relation
        \end{tabular}
        \caption{Constants}            
    \end{subfigure}    
    \caption{Relational logic}
    \label{fig:relational}
\end{figure}

The closure operators work only on binary relations and allow us to express reachability properties. Although not as powerful as the full recursion supported by \TLAplus, they are simpler to use, allow more succinct specifications, and are more amenable for automatic verification. The transitive closure operator is defined as follows.
\begin{displaymath}
    \ma{^} R \equiv \{ (a,b) \mid (a,b) \in R \vee \exists c \cdot (a,c) \in R \wedge (c,b) \in \ma{^} R \}
\end{displaymath}
Seeing binary relations as edges in a graph, $\ma{^} R$ denotes the relation containing all pairs $(a,b)$ such that there is a path from $a$ to $b$ through $R$-edges. For example, \a{Initiator.^adj} is the set of all nodes reachable from the initiator. The reflexive transitive operator also includes \a{iden}, the identity relation that maps each atom to itself.
\begin{displaymath}
    \ma{*} R \equiv \ma{^} R \mathbin{\ma{+}} \ma{iden}
\end{displaymath}

The assumptions about the network in the Echo protocol can be specified with relational logic as follows.
\begin{alloy}
fact ValidConfigurations {
    // no self loops
    no adj & iden
    // undirected graph
    adj = ~adj
    // all nodes reachable from initiator
    Node-Initiator in Initiator.^adj
}    
\end{alloy}
As we can see in this example, constraints in \Alloy\ tend to be specified in a very terse style. The relational logic operators enable a compositional (navigational) style, that usually requires less quantifiers, or no quantifiers at all as was the case here. This no-variables style is well-known in the algebra of programming~\cite{bird1997algebra} community, where it is known as the \emph{point-free} style of programming.

As mentioned above, we can ask the \Alloy\ \Analyzer\ to return instances satisfying our assumptions. This can be done with \a{run} commands. To ensure decidability all the commands must limit the size of the domain, by setting a \emph{scope} for all top-level signatures (the maximum number of atoms they can contain). The default scope is 3, but can be parametrized in each command. The \Analyzer\ also encodes by default a symmetry breaking mechanism that will try to exclude from the analysis all isomorphic instances up to renaming of atoms.
For example, to get instances of networks with up to 5 nodes we could execute the following command.
\begin{alloy}
run Example {} for 5    
\end{alloy}

\begin{figure}[t]
    \centering
    \begin{subfigure}[b]{0.45\linewidth}
        \centering
        \includegraphics[width=0.5\linewidth]{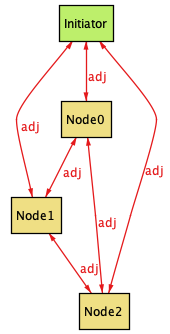}
        \caption{A complete network with 4 nodes}
        \label{fig:complete}
    \end{subfigure}
    \begin{subfigure}[b]{0.45\linewidth}
        \centering
        \includegraphics[width=0.5\linewidth]{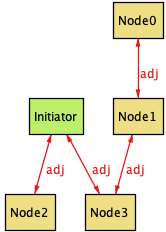}
        \caption{A line network with 5 nodes}
        \label{fig:line}
    \end{subfigure}
    \caption{Example network configurations}
    \label{fig:instances}
\end{figure}

By default, the \Analyzer\ shows instances graphically, with atoms being depicted as nodes and binary relations as edges between them. Atoms are automatically named according to the signatures they belong to. This visualization can be customized with themes, for example setting different colors and shapes for particular signatures. Figure~\ref{fig:instances} depicts two of the instances returned by the previous command, with the initiator customized to be shown in green. 

We can instruct the \Analyzer\ to search for specific instances. For example, the following command will directly return a complete network with 4 nodes (the one on Figure~\ref{fig:complete}).
\begin{alloy}
run CompleteNetwork {
    all n : Node | Node-n in n.adj
} for exactly 4 Node       
\end{alloy}
    
After validating our model with \a{run} commands, we can instruct the \Analyzer\ to check expected assertions, for example, to check that all networks satisfying \a{fact ValidConfigurations} are necessarily connected.
\begin{alloy}
assert Connected {
    all n : Node | Node-n in n.^adj
}
check Connected for 4    
\end{alloy}
In this case we checked this assertion for all networks with up to 4 nodes, and no counter-example was returned, as expected. Although not a full proof, the scoped analysis supported by \Alloy\ is still very useful in practice, since it exhaustively checks all instances up to a given size and most false assertions can be refuted with very small counter-examples, something known as the ``small scope hypothesis''.

\section{Behavioral design with \AlloyS}

Up to version 5, \Alloy\ only supported structural design -- all declared signatures and fields were immutable. It could be used for some behavioral design by resorting to an idiom that explicitly modelled traces, by declaring an explicit \a{State} signature over which a total order was imposed.  This signature was then added as an extra dimension to all ``mutable'' signatures and fields. This idiom was rather burdensome and error-prone, and only supported bounded model-checking of safety properties (exploring traces up to a given depth). A few attempts were made at extending \Alloy\ to natively support behavioral design, for example, by adding imperative-like programming constructs~\cite{near2010imperative} or dynamic logic~\cite{frias2005dynalloy}, but none ended up being part of the mainstream language, probably because they require substantial extensions to the language syntax and semantics.

Around 2015, together with colleagues at University of Porto and ONERA Toulouse, I also started working on an behavioral extension of \Alloy. Our main goal was to achieve a minimal extension that was aligned with the \Alloy\ philosophy. We ended up with a hybrid of \AlloyF\ and \TLA\ that we named \Electrum~\cite{MacedoEtAl16}. The only extensions to the \AlloyF\ language were the possibility of declaring mutable signatures and fields with the keyword \a{var},  support for the standard \LTL\ operators like \a{always} or \a{eventually} (but also past operators like \a{once}), and support for (a generalized version of) the prime operator of \TLA. With this combination, \Electrum\ allowed a system to be analyzed in the same style of \TLA, namely using the same logic to specify the behavior and the expected properties. Technically, the logic of \Electrum\ is not the \emph{Temporal Logic of Actions} but \emph{First-Order Linear Temporal Logic} (plus closures). In \Electrum\ prime can be used on any relational expression (not just on variables) and combined at will with the temporal operators, meaning there will be no guarantee of stutter invariance. Our goal was to give full freedom to the user when specifying the behavior, even if for most cases the typical pattern of specifying actions \emph{a la} \TLA\ will be followed. 

In the \Electrum\ \Analyzer~\cite{BrunelEtAl18} we added support for iteration over trace instances and counter-examples, support for bounded model-checking of arbitrary temporal formulas via translation to SAT, and support for complete model-checking by translation to the symbolic model-checkers \nusmv\ and \nuxmv. More details about the implementation of these features can be found in~\cite{macedo2022pardinus}. Perhaps due to its simplicity, \Electrum\ ended up being selected by the \Alloy\ board in 2022 to be the official \AlloyS\ version of the language.

\AlloyS\ excels precisely in the design of systems with complex configurations and data structures, that are better described with relations and relational logic. As we have seen in the previous section, it is rather trivial to specify the configurations of the Echo protocol in \Alloy. We will now show how the new features of \AlloyS\ allow us to also specify and verify its behavior.

Likewise in the \TLAplus\ specification, to model the dynamics of the Echo protocol in \AlloyS\ we can declare three mutable relations, each associating a node with: at most one \a{parent}; the set of nodes from which echos have already been \a{received}; and the set unprocessed messages in the respective \a{inbox}. The first two are binary relations. The latter is a ternary relation that associates nodes with the types of messages in the inbox and the respective senders. An abstract signature is also declared for denoting the types of messages, extended by two singleton signatures, one for each type. This ``abstract signature with singleton extensions'' pattern is the standard way of declaring an enumerated type in \Alloy. 
\begin{alloy}
sig Node {
    adj : set Node,
    var parent : lone Node,
    var received : set Node,
    var inbox : Node -> Type,
}
one sig Initiator extends Node {}

abstract sig Type {}
one sig Explorer, Echo extends Type {}     
\end{alloy}

To specify the behavior of the protocol we will use the same style of \TLAplus: a predicate without temporal operators to restrict the valid valuations in the initial state, conjoined with an invariant restricting the valid transitions, a disjunction of all possible events that can occur. 
\begin{alloy}
pred Init {
    // initiator is explored a priori
    parent = Initiator->Initiator
    // no received echos
    no received
    // initiator starts by sending explorers to all neighbors
    inbox = Initiator.adj->Initiator->Explorer
}
pred Next {
    some n : Node | receiveExplorer[n] or receiveEcho[n]
}
fact Spec {
    Init and always (stutter or Next)
}
\end{alloy}

As mentioned above, in \AlloyS\ stuttering is not mandatory, but in this case we opted to include it as a possible event in every state to make the specification similar to the one in \TLAplus. As usual in temporal logic, \AlloyS\ requires all traces to be infinite. Adding stuttering as a possible event at every state is a trivial way to ensure that. However, unlike with \TLAplus, we could restrict stuttering to only occur in particular states, for example, only when the protocol is finished. In the \a{Init} predicate we can see how \Alloy's relational logic simplifies the specification of constraints: to specify that the initiator starts by sending explorers to all neighbors, the initial state of the \a{inbox} is restricted to contain all triples that result from the cartesian product of \a{Initiator.adj} (the set of neighbors of the initiator) with the singletons \a{Initiator}
 (the sender of the messages) and \a{Explorer} (the type of messages).

 The \a{receiveExplorer} event can be specified as follows.
\begin{alloy}
pred receiveExplorer[n : Node] {
    some m : n.inbox.Explorer {
        no n.parent implies {
            parent' = parent + n->m
        } else {
            parent' = parent
        }
        no n.parent and some n.adj - m implies {
            inbox' = inbox - n->m->Explorer + (n.adj-m)->n->Explorer
        } else {
            inbox' = inbox - n->m->Explorer + m->n->Echo
        }
    }
    received' = received
}
\end{alloy}
Again we followed a style very similar to the one in the \TLAplus\ specification, namely specifying the full next-state value for each mutable relation. Note how relational logic again allows us to write some constraints more succinctly when compared to \TLAplus\ (in particular the specification of the new value of \a{inbox} in the case a node has to propagate an explorer to all its neighbors). However, with the \AlloyS\ \Analyzer\ we don't have the same restrictions imposed by \TLC\ when specifying events. In particular it is not mandatory to always define the value of every mutable relation using an equality or an inclusion test. For example, many \AlloyS\ users actually prefer to specify only the local effects in a relation and specify the frame conditions in a separate constraint. Instead of specifying the new value of \a{parent} using the restriction \a{parent' = parent + n->m}, that considers the global value of \a{parent} at once, we could have specified the local effect on the receiving node as \a{n.parent' = m} and separately specify the frame condition for all other nodes with the restriction \a{all x : Node - n | x.parent' = x.parent}. None of these restrictions would be accepted by \TLC.

\begin{figure}[t]
    \begin{subfigure}{0.5\linewidth}
        \centering
        \includegraphics[width=\linewidth]{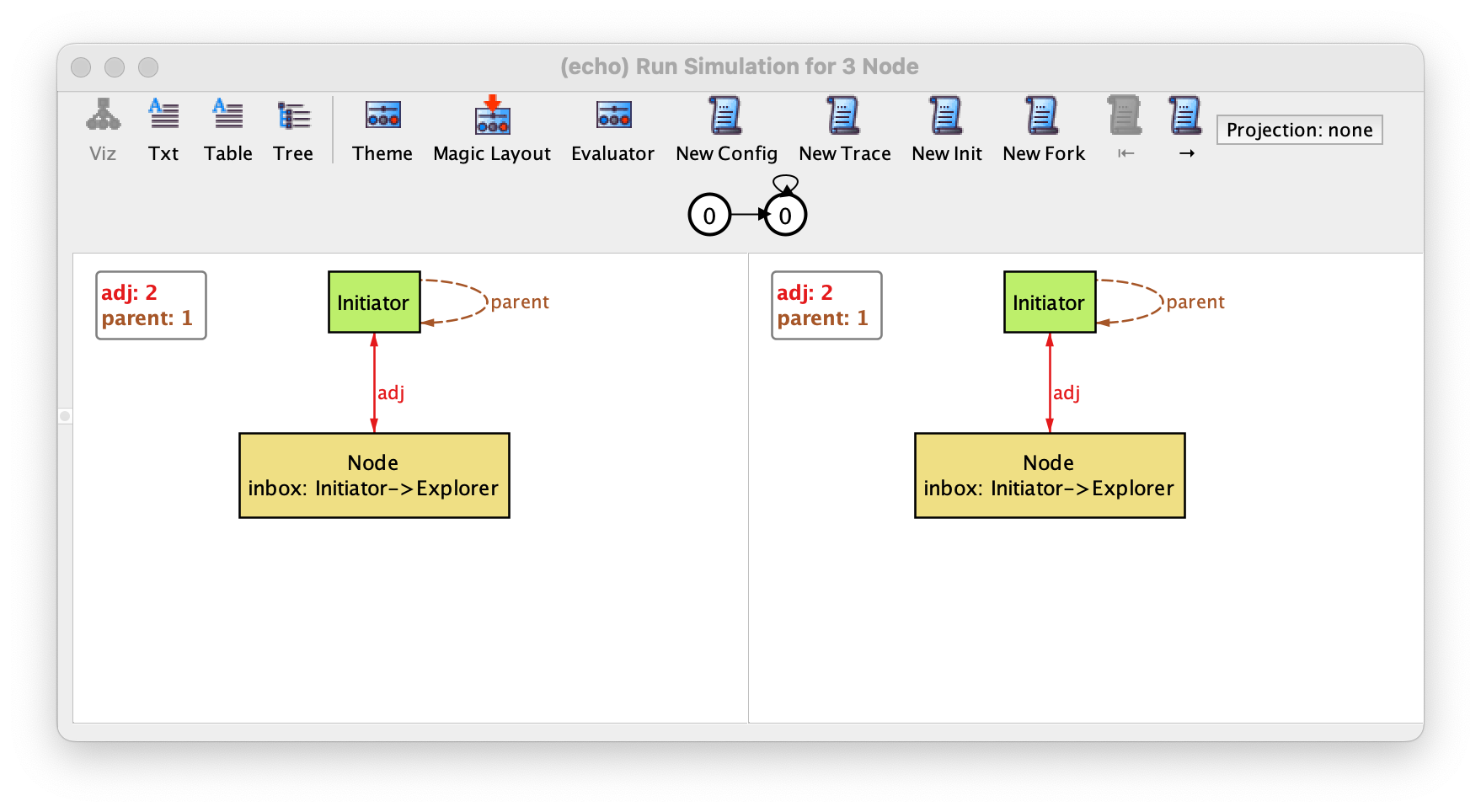}
        \caption{First returned instance}
        \label{fig:instance1}
    \end{subfigure}
    \begin{subfigure}{0.5\linewidth}
        \centering
        \includegraphics[width=\linewidth]{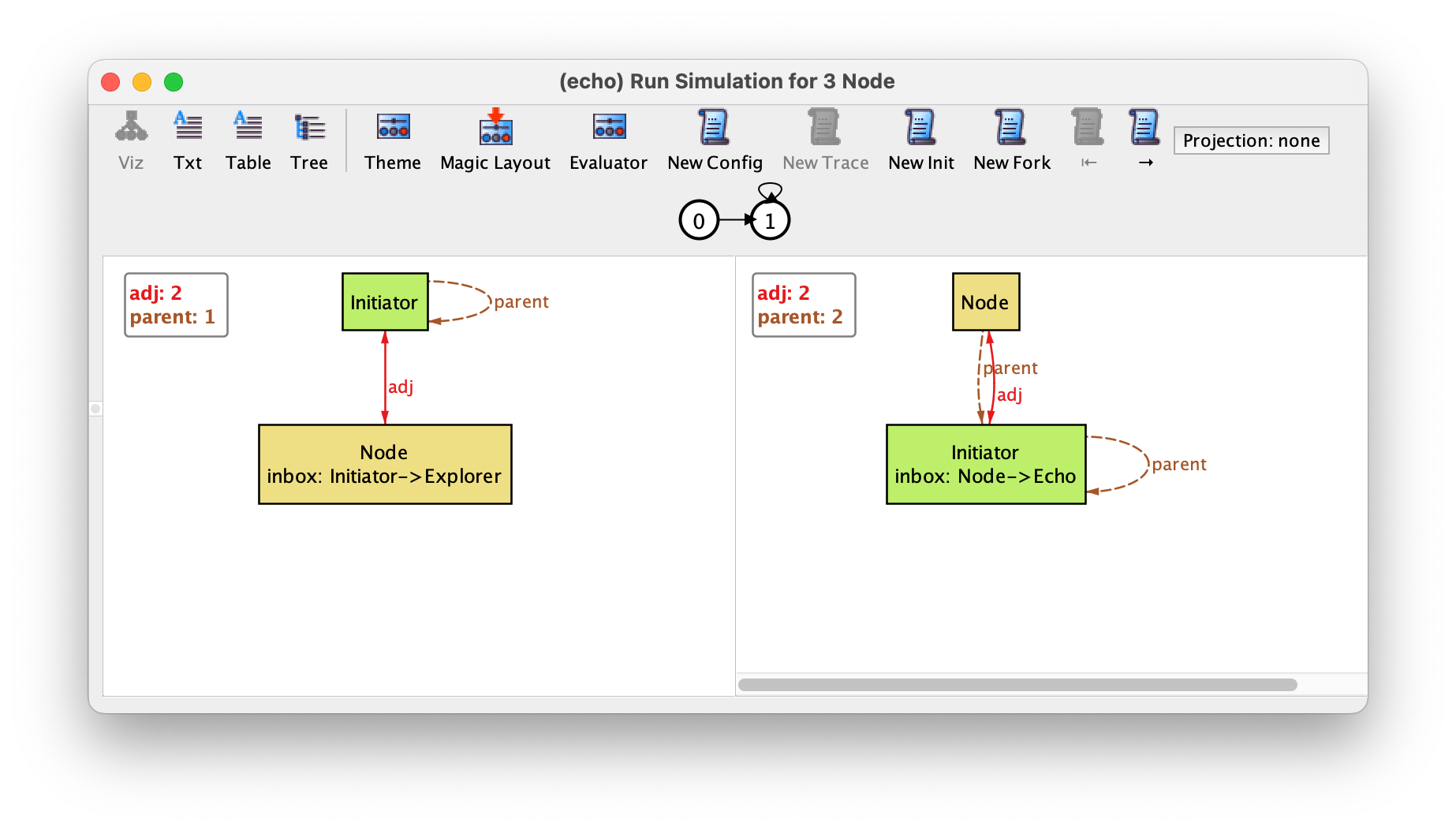}
        \caption{Result after \a{New Fork}}
        \label{fig:instance2}
    \end{subfigure}
    \begin{subfigure}{0.5\linewidth}
        \centering
        \includegraphics[width=\linewidth]{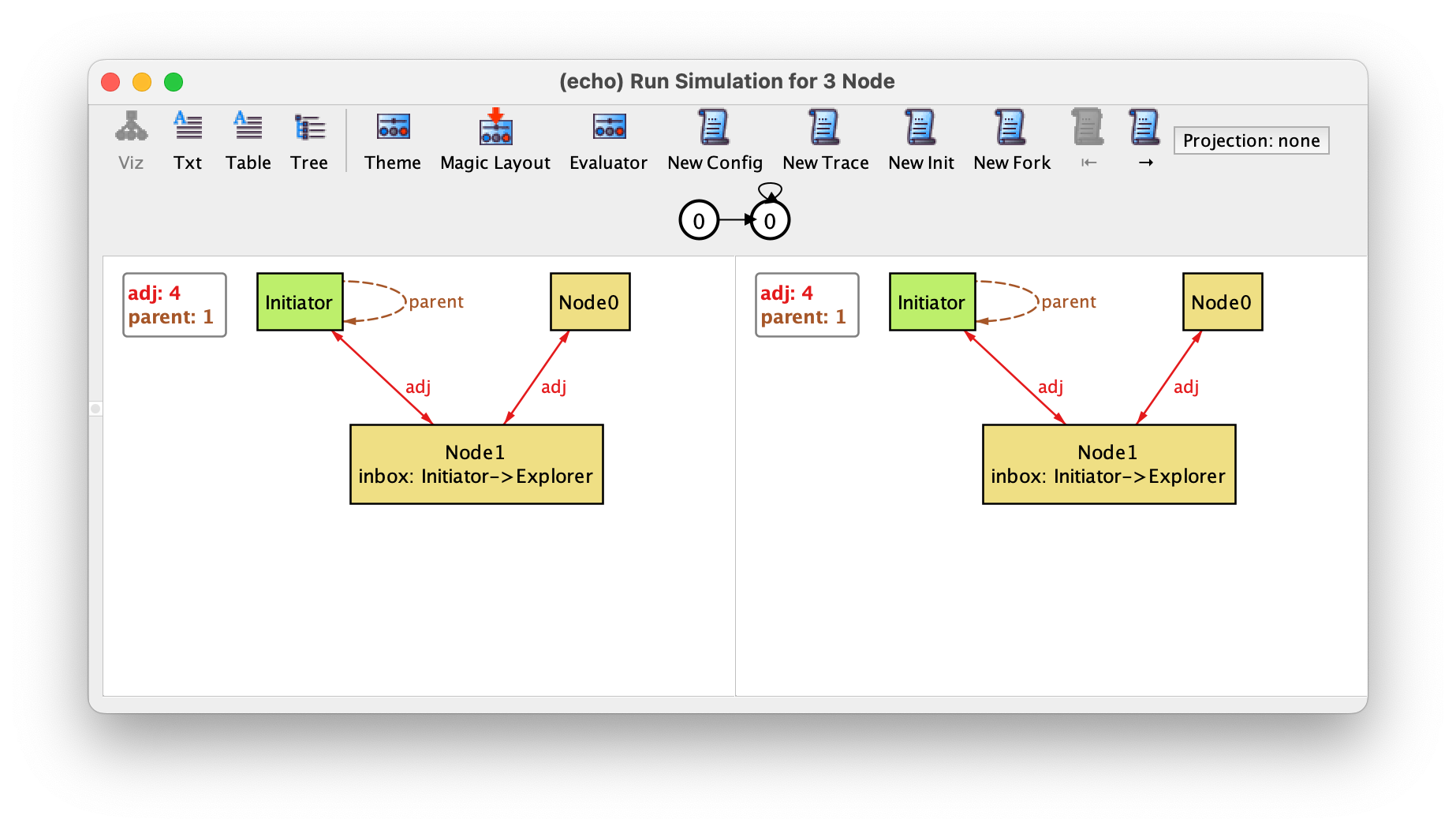}
        \caption{Result after \a{New Config}}
        \label{fig:instance3}
    \end{subfigure}
    \begin{subfigure}{0.5\linewidth}
        \centering
        \includegraphics[width=\linewidth]{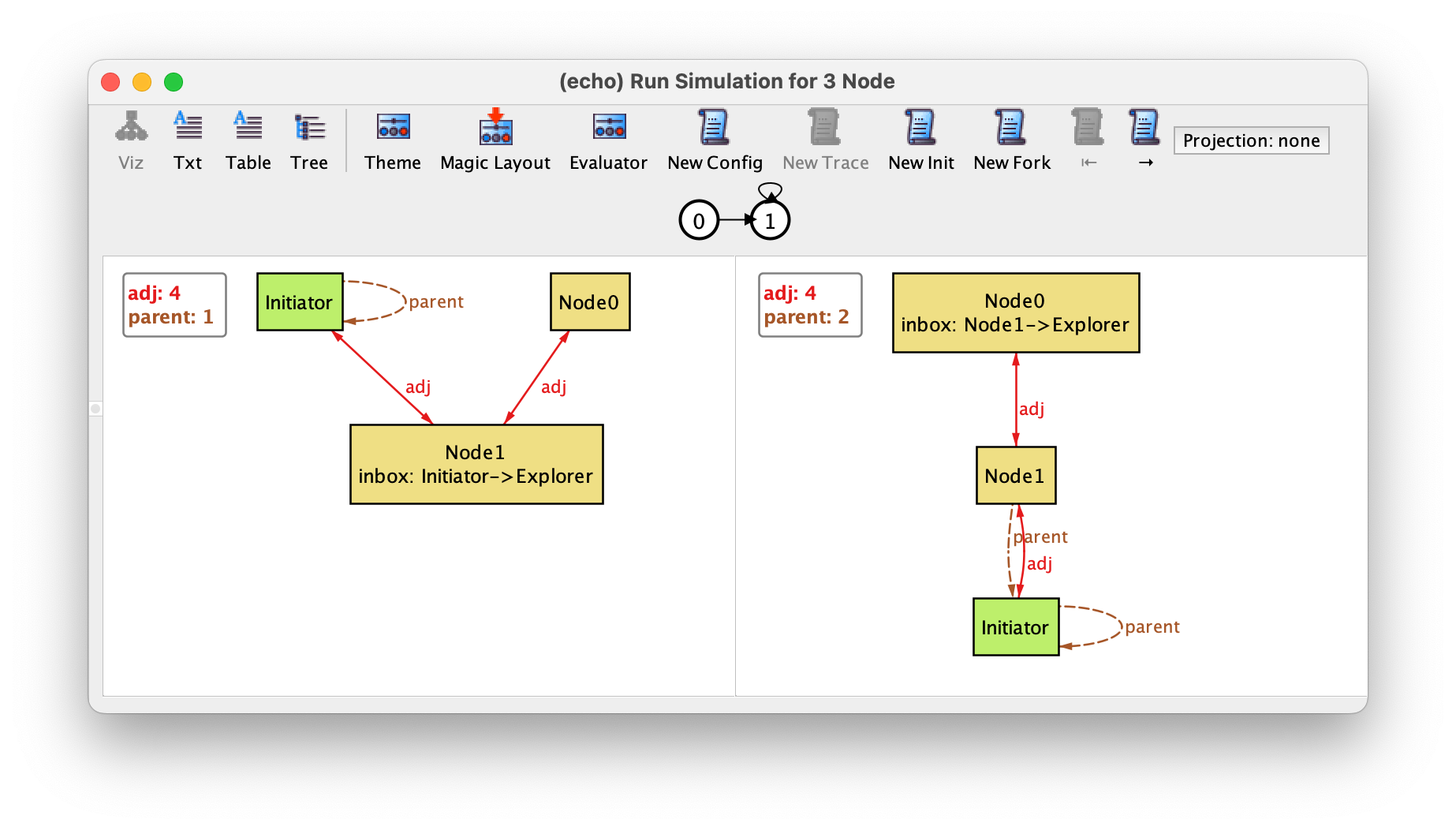}
        \caption{Result after \a{New Fork}}
        \label{fig:instance4}
    \end{subfigure}
    \caption{Interactive scenario exploration}
\end{figure}

Also, unlike with \TLC, we can directly ask for execution examples without having to specify false properties that produce the expected scenario. In particular, we can ask the \Analyzer\ to return an arbitrary scenario with the following command.
\begin{alloy}
run Simulation {}    
\end{alloy}
In this command the default scope of 3 is being used, so the returned scenarios will be for networks with up to 3 nodes. The first returned execution trace is shown in Figure~\ref{fig:instance1}, already customized with a theme that shows the values of the \a{inbox} and \a{received} as attributes of the nodes. As can be seen, the \AlloyS\ \Analyzer\ depicts one transition (or step) of the trace at a time, showing side-by-side in a split-panel the pre- and post-state to make it easier for the user to understand the event that occurred. On  top of this split-panel a depiction of the trace is also shown, identifying the different states and which segment of the trace is repeated indefinitely. The user can choose to focus on a particular transition by clicking directly any state or navigating forward and backward. In this case we can see that the first returned trace consists of an infinite sequence of stuttering steps in a network with 2 nodes, a valid execution of the protocol according to our specification.

The \AlloyS\ \Analyzer\ supports several trace iteration operations that allow the user to explore interactively different scenarios~\cite{fide19}. When no mutable relations are declared the \Analyzer\ defaults to the standard \AlloyF\ visualizer that only supported a single iteration operation (just asking for any different instance). In particular, we have the following operations for exploring traces: \a{New Config}, that returns an example trace for a different configuration (a different valuation to the immutable relations); \a{New Trace}, that returns any different trace for the same configuration; \a{New Init}, that returns a trace for the same configuration but with a different initial state; and \a{New Fork}, that returns a new trace for the same configuration that is the same up to the shown pre-state, but different afterwards. \a{New Fork} is particularly useful when exploring scenarios interactively, since it allows the user to see which different events are possible at each state. For example, in the first returned instance the user could ask for a \a{New Fork} to see a non-stuttering event. The resulting trace is shown in Figure~\ref{fig:instance2}, where the first event now corresponds to the non-initiator node receiving the explorer message, setting the initiator as its parent, and sending back an echo. Selecting \a{New Config} could yield the trace in Figure~\ref{fig:instance3}, where we now have a infinite stuttering trace for a line network with 3 nodes. Selecting \a{New Fork} on this trace returns the trace in Figure~\ref{fig:instance4}, where the middle node receives the explorer from the initiator and sends an explorer to the last node in the line. 

Using the different scenario exploration operations the user can easily validate the specification interactively in the early phases of the design. But it is also possible to ask directly for specific scenarios. For example, the following command asks directly for a trace where the protocol runs to completion in a complete network of 4 nodes.
\begin{alloy}
pred Finish {
    Initiator.received = Initiator.adj
}    
run MinimalRunComplete {
    all n : Node | Node-n in n.adj
    eventually Finish
} for exactly 4 Node, 19..19 steps        
\end{alloy}
By default, all analysis commands are run with the bounded model-checking engine, exploring traces up to depth 10 in the implicitly specified transition system. This default scope can be changed by setting the scope for \a{steps}. In the above command, since we know a complete run for this configuration takes 19 states, the scope of \a{steps} is set to be exactly that value to make the analysis run faster. The bounded model-checking analysis always tries to return the shortest instance or counter-example, so it will explore traces of increasing sizes in sequence, meaning that it can take a long time to get to a size where the desired instance is possible. Further restrictions can be imposed when searching for specific scenarios, for example, we can ask the \Analyzer\ to produce a trace where a particular sequence of events occurred, something not possible with \TLAplus.

To check the correctness of the protocol for all configurations with up to 4 nodes we can use the following command.
\begin{alloy}
pred SpanningTree {
    all n : Node - Initiator | Initiator in n.^parent
}    
assert Correctness {
    always (Finish implies SpanningTree)
} 
check Correctness for 4 but 1.. steps
\end{alloy}
Here we used the special scope \a{1.. steps} to trigger the complete model-checking analysis and verify all possible traces allowed by the specification (likewise to \TLC). With the symbolic model-checker \nuxmv\ selected as verification engine, this property is verified in a couple of seconds. For all configurations with up to 5 nodes it takes around 1m, way faster than the 13m that \TLC\ took to verify the same set of configurations. For all configurations with up to 6 nodes, the \AlloyS\ \Analyzer\ takes around 30m, while \TLC\ did not even finish enumerating all possible configurations after 1h. This efficiency is due to the sophisticated symbolic model-checking procedures implemented by \nuxmv\ to verify invariants, which are better suited for very non-deterministic systems such as this one (both in terms of different configurations and different interleavings).

To verify termination, we need to impose weak fairness. Unlike with \TLAplus, in \AlloyS\ there is no special keyword for that and the user must explicitly encode the required weak fairness.
\begin{alloy}
pred Fairness {
    eventually always some inbox implies always eventually Next
}    
assert Termination {
    Fairness implies eventually Finish
}
check Termination for 4 but 1.. steps    
\end{alloy}

Unfortunately, the symbolic model-checking engines currently supported by \AlloyS\ are all quite slow at verifying this property. For networks with up to 4 nodes the best timing was around 20m for complete model-checking, way slower than \TLC\ for the same scope. For bigger scopes the verification timed out at 1h.

\section{Final remarks}

As we have seen, the formal design of software with complex configurations is not only desirable but already possible and cost-effective with the presented languages and tools. 
The semantics of both \Alloy\ and \TLAplus\ are based on very simple mathematical concepts that most software engineers should have learnt in their undergrad, and both provide automatic analysis tools that require little expert input, so I believe they have the potential to be used at large.

Although the analysis of multiple configurations is not natively supported by \TLAplus, it is rather trivial to adapt a specification that checks a single configuration to one that checks multiple configurations at once. The main problem is that the \TLAplus\ logic and the explicit model-checking technique of \TLC\ are not best suited to specify and enumerate configurations with complex structure and constraints. For designing software with simple configurations, such as concurrent algorithms where only the number of nodes varies, \TLAplus\ and \TLC\ should be an excellent choice, even if \TLC\ limitations can sometimes be frustrating.

\Alloy\ on the other hand excels at describing and analyzing complex configurations, and the symbolic model-checking techniques implemented in the \AlloyS\ \Analyzer\ are already efficient enough to verify safety properties for many complex configurations at once. The visualization and scenario exploration operations implemented in the \Analyzer\ also make \Alloy\ substantially better for the validation task in these contexts. 
In recent years we successfully applied \Electrum\ and \AlloyS\ in the analysis of some larger case studies of software with complex configurations, for example the Hybrid ERTMS/ETCS Level 3 railway signaling standard~\cite{CunhaM20} and interactive systems described with task models~\cite{taskmodels}.

What about other formal software design frameworks? As we have mentioned in the introduction, many are heavyweight formal methods, for example theorem provers, requiring a lot of expertise input and thus only viable to apply in the most safety-critical applications. But even in those applications, using frameworks like \TLAplus\ or \Alloy\ can help with debugging and validation in the early phases of the design, making sure one only attempts to do the full proofs after having a lot of confidence the design is correct. As for lightweight formal methods, the most popular model-checkers -- for example \nusmv, \nuxmv, or \spin\ -- are too low-level to be used directly to describe software with complex configurations, being better suited to be used as backend analysis tools for higher-level languages, as is the case with \AlloyS. Among the higher-level formal specification languages, perhaps the tools in the \B\ method~\cite{abrial1996b} ecosystem are the ones better suited to compete with \TLAplus\ and \Alloy, in particular the \ProB\ tool~\cite{leuschel2008prob}. This tool already implements some of the features of the \Alloy\ \Analyzer, namely simulation and graphical depiction of counter-examples, and offers interfaces to both \TLAplus\ and \Alloy. Unfortunately, its specification language is far from minimal
and, in my opinion, not as ideal for abstract specification as that of \Alloy\ or \TLAplus.

There is still a lot of room for improvement, though. None of the tools was good enough to check liveness properties for reasonable network sizes, so further research on improving the scalability of the model-checking techniques is necessary. Also, one should consider extending these languages to better support the quantitative requirements that are becoming pervasive in modern software, for example in machine learning systems or security protocols. For \Alloy\ we have already done some preliminary steps in that direction~\cite{silva2022quantitative}.

\bibliography{references}

\end{document}